\documentclass{aa}
\usepackage{txfonts}
\usepackage{amsmath}
\usepackage{amsfonts}
\usepackage{amssymb}
\usepackage{makeidx}
\usepackage{graphicx}
\usepackage{url}
\usepackage{comment}
\usepackage{multirow}
\usepackage{xcolor}
\usepackage{aas_macros} 

\begin{document}

\title{RATAN-600 multi-frequency data for the BL Lac objects \thanks{The catalogue is presented in interactive form and available at Special Astrophysical Observatory of the Russian Academy of Sciences website \protect\url{http://www.sao.ru/blcat/}}}
 \author{M.G.~Mingaliev
          \inst{1,2}\thanks{marat@sao.ru}
          \and
	  Yu.V.~Sotnikova\inst{1}\thanks{sjv@sao.ru}
          \and
	  R.Yu.~Udovitskiy\inst{1}
         \and
	  T.V.~Mufakharov\inst{1}
         \and
	  E.~Nieppola\inst{3,4}
	  \and
	  A.K.~Erkenov\inst{1}
          }

   \institute{Special Astrophysical Observatory of RAS, Nizhnij Arkhyz, 369167 Russia\\
              \email{marat@sao.ru}
         \and
             Kazan Federal University, 18 Kremlyovskaya St., Kazan, 420008, Russia
	  \and
	     Finnish Centre of Astronomy with ESO (FINCA), University of Turku, V{\"a}is{\"a}l{\"a}ntie 20, FI-21500 Piikki{\"o}, Finland
          \and
             Aalto University Mets{\"a}hovi Radio Observatory, Mets{\"a}hovintie 114, 02540 Kylm{\"a}l{\"a}, Finland
             }

   \date{Received August 15, 2014; accepted August 16, 2014}

 
  \abstract
{}
{We present a new catalogue of the RATAN-600 multi-frequency
measurements for BL Lac objects.
The purpose of this catalogue is to compile the BL Lac multi-frequency data
that is acquired with the RATAN-600 simultaneously at several frequencies.
The BL Lac objects emit a strongly variable and polarized non-thermal radiation
across the entire electromagnetic spectrum
from radio to $\gamma$ rays and represent about 1{\%} of known AGNs.
They belong to the blazar population and differ from other blazars' featureless optical
spectrum, which sometimes have absorption lines, or have
weak and narrow emission lines. One of the most effective ways of studying
the physics of BL Lacs is the use of simultaneous multi-frequency data.}
{The multi-frequency broadband radio spectrum was obtained
simultaneously with an accuracy of up to 1-2 minutes
for four to six frequencies: 1.1, 2.3, 4.8, 7.7, 11.2, and 21.7 GHz.
The catalogue is based on the RATAN-600 observations and
on the data from:
equatorial coordinate and redshift,
R-band magnitude, synchrotron peak frequency,
SED classes and object type, literature.}
{The present version of the catalogue contains RATAN-600 flux densities measurements over nine years (2006--2014),
radio spectra at different epochs, and their parameters
for more than 300 BL Lacs objects and candidates.
The BL Lacs list is constantly updated with new observational data of RATAN-600.}
   {}


\maketitle

\section{Introduction}
The objects of BL Lacertae-type (BL Lac or BLO) are a relatively rare class of active galactic
nuclei (AGN); they represent about 1\% of known AGN \citep{2008AJ....135.2453P}.
They emit electromagnetic radiation across the entire spectrum from radio to
$\gamma$ rays \citep{1995PASP..107..803U}. The BL Lacs are characterised by the absent
or weak emission lines in the optical spectrum, the strong and rapid variations
in intensity, and polarization
\citep{1972ApJ...175L...7S,1994VA.....38...29K,1995PASP..107..803U,1998MNRAS.301..985S}.
These objects are known to be radio-loud; it is usually explained with the presence
of relativistic jet, which is disposed at a small angle to the line of sight of the observer
($\theta$ $<$ $20^{\circ}$). Thereby, relativistic effects play a major role in
the formation of observational properties of BL Lacs \citep{1978bllo.conf..328B}.
Systematic studies of BL Lacs objects were historically limited to the shallow surveys and small
incomplete observation samples, which had a major systemic effect on the results
\citep{2008AJ....135.2453P,2009A&A...495..691M}. Unlike other types of AGNs, BLO objects
were detected in radio and X-ray surveys \citep{1990ApJ..348..141S},
which are followed by separation of the class into two large groups, XBLs (X-ray
selected BLOs) and RBLs (radio-selected BLOs). Various observational
properties (intensity, polarization variability, and luminosity) suggest
that these objects may have a different nature. Radio-selected objects are
on average more luminous, variable, and polarized, and their structure is
more core-dominated than that of X-ray-selected objects
\citep{1993ApJ...406..430P,1994ApJ...428..130J,1991ApJ...380...49M,1992A&A...256..399P}.
In the past ten years, the Sloan Digital Sky Survey (SDSS) has presented an opportunity to
search for BL Lac objects by exploiting their featureless optical continuum one of their defining characteristics.
The SDSS surveys have been used to produce a catalogue of
candidate BL Lac objects by finding quasi-featureless spectra within their huge spectroscopic
databases \citep{2005AJ....129.2542C}. This type of BLOs could be classified as optically-selected (OBLs).

The spectral energy distribution (SED)
of the blazars is characterised by two broad
features. The feature peaking at lower energy is generally explained
in terms of synchrotron emission and the second feature, which peaks at
higher energies, is likely due to the Inverse Compton scattering.
We adopt BL~Lac classification from \citet{2006A&A...445..441N}.
According to the synchrotron peak frequency ($\nu_{peak}^S$), the BL~Lacs can be
separated to three subclasses: HSPs, ISPs and LSPs (HSPs, which are
BL Lacs having $\nu_{peak}^S$ higher than $\sim 10^{16.5} Hz$, LSPs, which have a
$\nu_{peak}^S$ lower than $\sim 10^{14.5} Hz$ and ISPs, which have an intermediate
synchrotron peak). In general, HSPs are
very weak at the radio band \citep{2007AJ....133.1947N}
and often correspond to X-ray-selected
objects, while LSPs are related to radio-selected objects.

Most of BL~Lac samples are based on the cross-correlation of
existing radio and X-ray catalogues \citep{2007A&A...472..699T}.
``Roma-BZCAT: a multi-frequency catalogue of blazars''
\citep{2009A&A...495..691M} is the most popular catalogue
of the blazars for today. It is based on multi-frequency
data from various surveys and includes 1221 objects of BL Lac
type (for Edition 4.1.1). For this catalogue and some
others (e.g., \citet{2008AJ....135.2453P}), the
multi-frequency and simultaneous observation data covering
long time intervals are presented for a relatively small
number of objects.
The number of new BL~Lac candidates is growing
constantly, and this is due to the large number of the new sky
missions in different frequency ranges.
Many of BL~Lacs have only been observed once or twice,
and their nature is still not clear.

Long-term monitoring is also important because of the variability
present in BL Lacs. Sometimes, some sources show changes in the flux
density of up to hundreds of a percent, but they are also observed to remain in the
quiescent state for years (see, e.g., \citet{1995ARA&A..33..163W}).
A detailed, versatile, and prolonged research of BL Lacs is essential
for understanding the AGN phenomenon, which are unusual properties of these
objects and unique physical processes occurring there
(emission mechanisms in details, origin of the emitted photons, jets structure, etc.).
That is why it could be useful and important to collect
multi-frequency data, which is measured quasi-simultaneously at
one instrument for a long period of time.

Hereinafter, we refer to the catalogue as ``BLcat''.
Observational data and radio properties corresponding to each object from
our catalogue are available in \citet{2001A&A...370...78M,2007ARep...51..343M,2012A&A...544A..25M}.
In this paper, we present the common organisation of the electronic catalogue of the BL Lac objects.
The paper is organized as follows. In Section 2, we describe RATAN-600 observations and
characteristics of the receivers. In Section 3, we present the general organisation
of the catalogue.
The sample properties are given in Section 4, where we describe the flux densities,
radio continuum spectra, variability, spectral indices, and also SED classes.
The detail analysis of radio properties of objects from BLcat are
presented in Mingaliev et al. 2014 (in preparation).

\section{RATAN-600 observations}

The observations were carried out with the RATAN-600 radio telescope during 2006-2014.
This list originally included 108 objects from the Mets{\"a}hovi BL Lac sample \citep{2006A&A...445..441N}.
Systematic monitoring was carried out during the period of 2006--2008 for these sources. In
2009--2011, we have observed only a part of sample, due to constraints on
observing time.
We considerably increased the size of (by more than a factor of three)
our BL~Lac sample since 2012, and thus,
most of the newly added sources have been observed only a couple of times.
Since 2012, we added sources with the flux density greater than
400~mJy at 1.4 GHz; since 2014,
we began adding BL Lacs with a flux density greater than 100~mJy.

Observations were carried out with two sets of radiometers.
The first ones (noted as ``1'' in Table~\ref{tab:radiometers}) at the
wavelength 1.38, 2.7, 3.9, 6.2 cm (cryogenically cooled), and
1.1 and 2.3 cm (uncooled). The second set is uncooled (noted as ``2'') at the
wavelength 1.38, 2.7, and 6.2 cm.
All the radiometers were designed as the ``direct amplification Dicke type''
receivers. The 6.2 cm cooled radiometer is a noise added radiometer (NAR),
while other radiometers are designed according to the beam-switching scheme.
For all continuum radiometers, we use the data acquisition and
controling system, as described by \citet{2011AstBu..66..109T}.

Observations were made in a transit mode at the north and south
sectors of the antenna,
and the multi-frequency broadband radio spectrum was obtained by observing simultaneously
with an accuracy of up to 1-2 minutes \citep{1993IAPM...35....7P}.
An angular resolution in this mode of observations depends on a declination of a source being observed.
The FWHM in right ascension (RA) is given in Table~\ref{tab:radiometers};
an angular resolution in declination is three to five times worse than in RA.
The experimental data were processed using the modules of the FADPS
(Flexible Astronomical Data Processing System) standard reduction
package by \citet{1997ASPC..125...46V}.
This is a data reduction system for broadband continuum
radiometers of the RATAN-600. The data processing technique is
described in \citet{2012A&A...544A..25M}.
The following twelve flux density calibrators (standard and
RATAN's traditional ones) were used to
get the coefficients of antenna elevation: 3C48,
3C138, 3C147, 3C161, 3C286, 3C295, 3C309.1, NGC7027,
J0237$-$23, J1154$-$35, J1347$+$12, and J0410$+$76.
Measurements of some calibrators were corrected for angular
size and linear polarization, following the data summarized in \citet{1994A&A...284..331O}
and \citet{1980A&AS...39..379T}, respectively.

The detection limit for the RATAN-600 single sector
is approximately 8~mJy (integration time is about 3 sec) under good conditions
at the frequency of 4.8 GHz and at an average antenna elevation ($\delta$$\sim$$42^{\circ}$).
At other frequencies, the detection limits for two radiometer systems
are presented in the Table~\ref{tab:radiometers}.
These values depend on the atmospheric extinction instability and the effective
area at the antenna elevation $H$ (from $10^{\circ}$ up to $90^{\circ}$ above the horizont)
at the corresponding frequencies.

\begin{small}
\begin{table}
\caption{RATAN-600 continuum radiometers.}
\label{tab:radiometers}
\centering
\begin{tabular}{rrllccr}
\hline\hline
\multicolumn{2}{c}{$f_{0}$}  & \multicolumn{2}{c}{$\Delta$$f_{0}$} & \multicolumn{2}{c}{$\Delta$$F$}  & FWHM       \\
\multicolumn{2}{c}{(GHz)}    & \multicolumn{2}{c}{(GHz)}           & \multicolumn{2}{c}{(mJy/beam)}   & (arcsec) \\
\hline
      1     &     2           &   1        &  2              &  1   & 2 &         \\
\hline
\hline
$21.7$ & $21.7$ & $2.5$  & $2.5$  &  $70$  & $88$  &   11 \\
$11.2$ & $11.2$ & $1.4$  & $1.0$  &  $20$  & $20$  &   16 \\
$7.7$  & ...    & $1.0$  & ...    &  $25$  & ...   &   22 \\
$4.8$  & $4.8$  & $0.9$  & $0.8$  &  $8$   & $11$  &   36 \\
$2.3$  & ...    & $0.4$  & ...    &  $30$  & ...   &   80 \\
$1.1$  & ...    & $0.12$ & ...    &  $160$ & ...   &  170 \\
\hline
\hline
\end{tabular}

\tablefoot{
Column designation:
Col.~1~-- central frequency,
Col.~2~-- bandwidth,
Col.~3~-- flux density detection limit per beam, and
Col.~4~-- angular resolution (FWHM in RA).}
\end{table}
\end{small}

Due to the observational conditions and the natural limits set by the source
flux levels, the number of data epochs vary from tens to one
at the beginning of 2014.
Each set contains averaged measurements from three to ten transit scans.

\section{General organisation of the catalogue}

The current version of the catalogue includes more than 300
blazars, most of them are BL Lac objects and candidates.
The main part of the catalogue is
multi-frequency measurements, which are obtained quasi-simultaneously
with one instrument.
The list of objects with their general parameters and RATAN-600 data
are presented in an online form.
The main table displays the number of RATAN-600 observations, equatorial coordinates, names,
redshifts, magnitudes in R-band,
average flux densities $F_{4.8GHz}$ (from RATAN-600 data), synchrotron peak frequencies,
SED classes (LSP, ISP, HSP), blazar (BL~Lac, BL~Lac candidate or Blazar uncertain type),
selected (XBL, OBL, RBL) types, and ADS references.
The second part of catalogue consists of RATAN-600 measurements of objects at four to six
frequencies in the period from 2006 to 2014.

\begin{figure*}
\centerline{\includegraphics[width= 0.9\paperwidth]{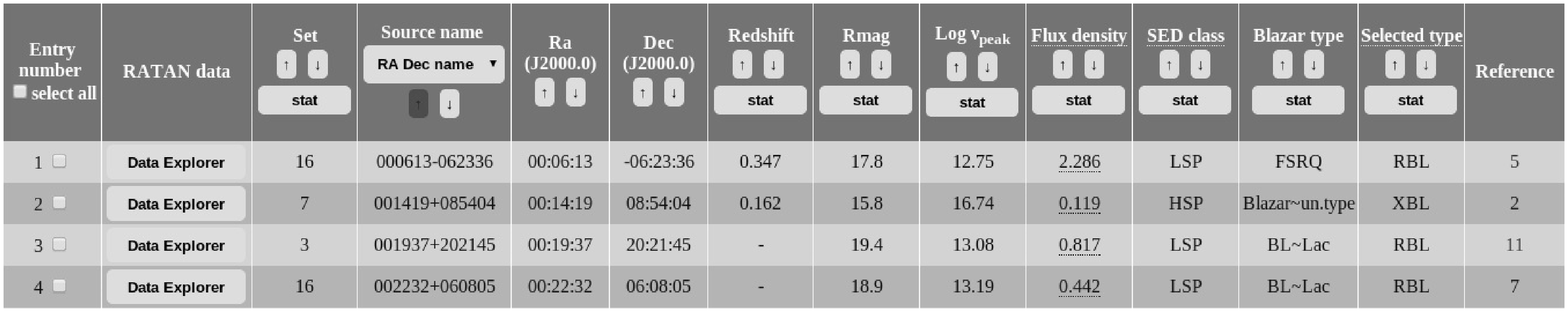}} 
\caption{BLcat main page screenshot.} 
\label{fig:scr}
\end{figure*}

\begin{figure}
\centerline{\includegraphics[width=65mm]{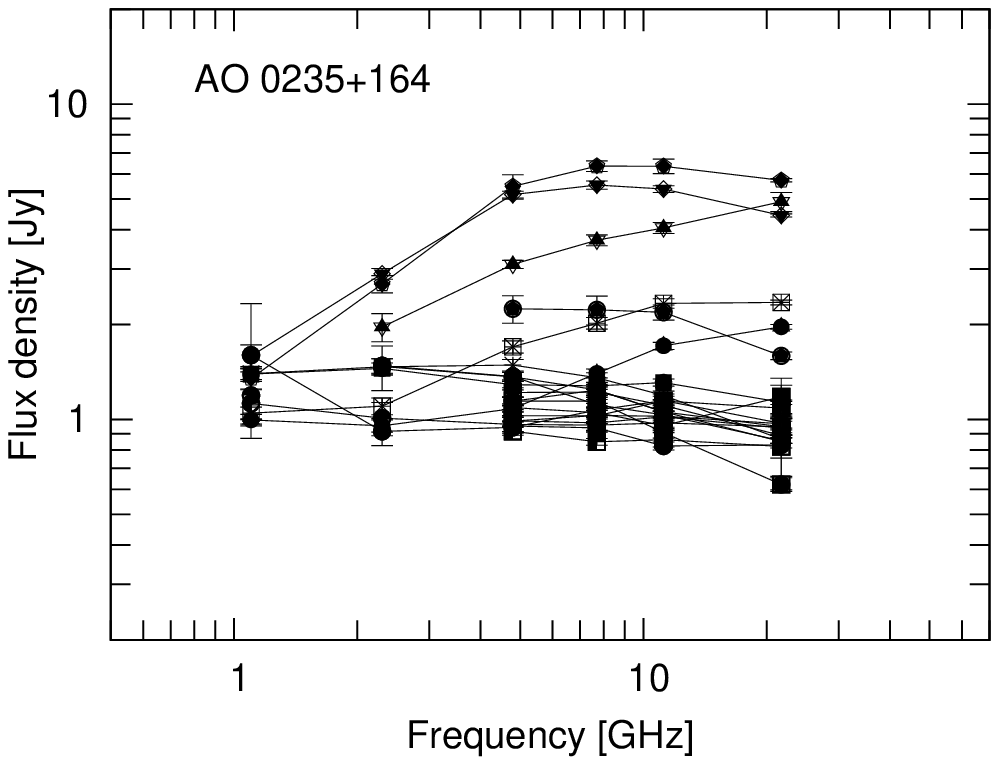}}
\centerline{\includegraphics[width=65mm]{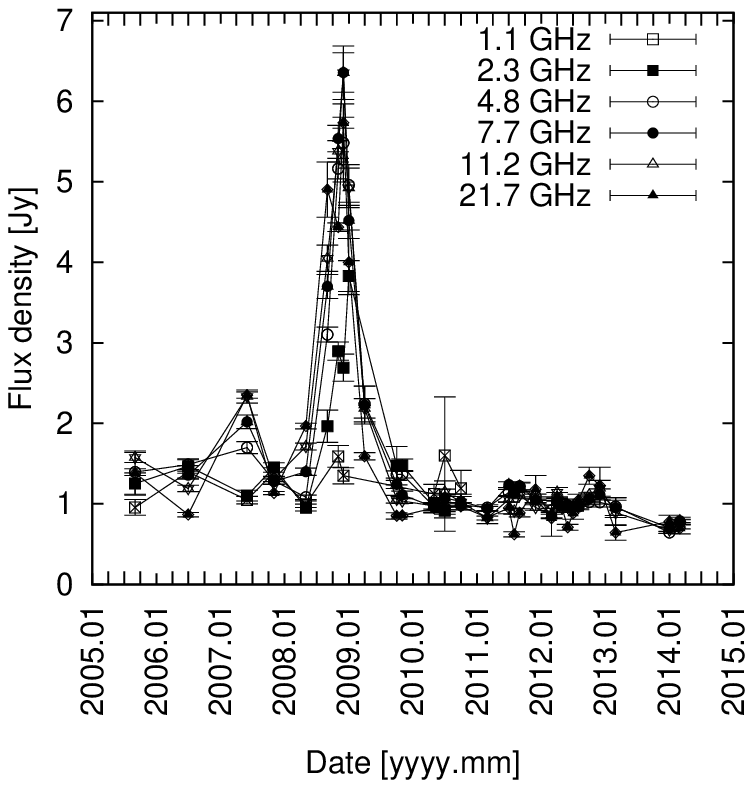}}
\caption{The multi-frequency radio spectrum and light curves
of the BL Lac object $AO 0235+164$
derived from the catalogue. There are 31 observational epochs for
this object from 2005.09 up to 2014.03.
Observational epochs are displayed in the electronic version of the catalogue automatically.}
\label{fig:example}
\end{figure}

The object list (Fig. \ref{fig:scr}) is available in electronic form
in different formats in the ``Export the main Table and RATAN-600 data''
button. The RATAN-600 data are available in the form of
radio continuum spectra and light curves (Fig.~\ref{fig:example}),
flux density tables with the date of observation
in the ``Data Explorer'' or ``Export the main Table and RATAN-600 data'' button.

For six sources (two ISP and four HSP), we have no
observational data, because they are too faint (tens of mJy)
to detect their flux density.
With the tool ``Export the main Table and RATAN-600 data'',
one can extract fluxes, date of observation, spectral indicies
calculated at different frequency intervals, and
download as a \textit{CSV} format text file.
The spectral index describes the slope of the spectrum,
and it is one of the important characteristics of
the radio spectrum.
We calculated spectral indices using the formula:
\begin{equation}
\label{sp:index}
\alpha=\frac{\log S_{2}-\log S_{1}}{\log{\nu}_{2}-\log{\nu}_{1}},
\end{equation}
where $S{_1}$ and $S{_2}$ are the flux densities at the frequencies $\nu{_1}$ and
$\nu{_2}$, respectively.

To characterize the variability properties of the sources
at the various frequencies, we have computed the variability index.
The variability index was calculated using the formula, as adopted by \citet{1992ApJ...399...16A}:
\begin{equation}
\label{variability}
Var_{S}=\frac{(S_{i}-\sigma_{i})_{max}-(S_{i}+\sigma_{i})_{min}}
{(S_{i}-\sigma_{i})_{max}+(S_{i}+\sigma_{i})_{min}},
\end{equation}
where $S_{max}$ and $S_{min}$ are the maximum and minimum value of the flux density
at all epochs of observations; $\sigma_{S_{max}}$ and $\sigma_{S_{min}}$
are their RMS errors.

\section{The sample properties}
\subsection{Classification}

The main table presents the classification of blazars by \citet{2009A&A...495..691M},
which is based on the spectral properties.
The BL~Lac objects have a featureless optical spectrum, or only have absorption lines and
weak narrow emission lines (rest-frame equivalent width, EW$<$5\AA) \citep{1995ApJ...444..567P}.
Flat-spectrum radio quasars (FSRQ)
have broad emission lines in their spectrum, and a property
that is as strong as non-thermal emission from radio waves to the $\gamma$-rays,
which is a presence of evidence for relativistic beaming. Blazars of an uncertain type
(Blazar~un.type) are the sources with peculiar characteristics and that show blazar activity.
There are seven FSRQs in the list. This because
these objects were classified as a BL~Lac initially in the Veron-Cetty and Veron
BL Lac Catalogue (Veron-Cetty \& Veron 2000),
and they were observed with RATAN-600 within the BL~Lacs monitoring program.

\begin{table}
\caption{The sample subpopulation classes.}
\label{tab:subpopulation}
\centering
\begin{tabular}{lll}
\hline\hline
Designation criterion & Class & Number\\
\hline\hline
\multirow{4}{*}{Optical spectrum} & BL~Lac & 220\\
		 & BL~Lac~cand. & 43\\
		 & Blazar~un.type & 36\\
		 & FSRQ & 7\\
\hline
\multirow{3}{*}{Selection method} & RBL & 248\\
		 & XBL & 56\\
		 & OBL & 2\\
\hline
\multirow{3}{*}{SED type} & LSP & 141\\
		 & ISP & 102\\
		 & HSP & 63\\
\hline
\hline
\end{tabular}
\end{table}

The majority of sources are BL Lac objects and candidates;
$\sim$10{\%} are uncertain Blazar-type objects and only seven are FSRQ type sources.
The sample subpopulations are presented in Table~\ref{tab:subpopulation}.
The abbreviation RBL/OBL/XBL corresponds to the frequency range, where the
object was selected. There are two OBL-type sources in the list for today
(US1889 and SDSSJ16581+6150). Most of the BL Lacs are RBL-type sources (248)
and 56 are XBL-type.

The redshift values were taken from Roma-BZCAT and NASA/IPAC NED database.
Redshift values are known for the most of sources of the catalogue (241), and
the peak of the redshift distribution occurs at $\sim$ 0.25.


\subsection{Flux densities and radio continuum spectra}

The flux densities of $\sim$300 sources are given in the table
form of the catalogue. The flux density uncertainties for
the corresponding frequencies include the uncertainty in the
source's antenna temperature and calibration curve.
Most objects have the data at 7.7 and 4.8 GHz.
The absence of data for some sources at some frequencies
is a result data exclusion because of the partial
resolution of a source, a source that is too weak
to be measured reliably, a strong influence of man-made
interference in the decimeter, and strong interferences
from the geostationary satellites at 11.2 GHz
(in the declinations between $-10^{\circ}$ and $0^{\circ}$).
Nevertheless, in some cases, the data was not excluded in spite of
the increase in errors. The values of the standard error of
fluxes for the most sources are: 5-20{\%} for 11.2, 7.7,
and 4.8 GHz; and 10-35{\%} for 2.3, 1.0, and 21.7 GHz.

Some sources in this database are extremely faint at radio frequencies,
and this is the first time that their radio spectra have been measured reliably.
The flux densities and spectra are studied in more detail in Mingaliev et al. 2014.

\subsection{The synchrotron peak frequency and SED class}

The synchrotron peak frequencies for some sources were calculated from
a parabolic fit to their SEDs \citep{2006A&A...445..441N},
and for the most objects, we used
``SED Builder'' tool\footnote{\url{http://tools.asdc.asi.it/SED}}.
Those $\log$~$\nu_{peak}^S$ values marked with ``*''.
The peak of the $\log$~$\nu_{peak}^S$ distribution occurs at $\sim$13.6 Hz for the current sample.
We note that some objects do not have enough data points for a
reliable estimations of $\log$~$\nu_{peak}^S$.
In these cases, the synchrotron peak frequency value depends on the method of
calculation (polynomial degree, for example).
If there are only data points in the low frequency range (radio domain),
the peak shifts to the low frequency region (see, for example, $MS~1133.7$+$1618$).
Any new observations at high
frequencies (optical and ultraviolet) improve the accuracy of the peak value.
Therefore, the presented SED types are not complete and depend
on the new observational data.
The LSP class of BL Lac dominates in the current version of the catalogue
(141 sources),
since we initially observed radio loud sources more often than others.
There are 102 objects with an intermediate synchrotron peak (ISP)
and 63 objects with a high frequency of synchrotron peak (HSP).

\section{Conclusion}

In this paper, we describe a practical tool, based on
the BL Lac observations with the RATAN-600 radio telescope for researchers. This catalogue
is an useful object list useful in the research of BL Lacertae populations,
or statistical studies of radio properties and evolution.
The main goal of this catalogue is to remedy the lack of data for
these objects in the radio domain.
The ``BLcat'', which is a collection of the multi-frequency (1--21.7 GHz) radio band measurements,
are obtained quasi-simultaneously with a single instrument over the course of several years. All the data
are presented online in a convenient form with interactive features.
Now, everyone can quickly have a look to the behaviour of a certain BL Lac object
from our list at the radio band, which includes examining spectra evolution or light curves;
some basic characteristics of the objects are also presented. The present
version includes more than 300 objects. The majority of them are BL Lac objects
and candidates, which about 10{\%} are Blazar uncertain type objects.

The online catalogue is updated regularly and filled with new
and previous observational data from the RATAN-600
radio telescope. The catalogue is available from the
Special Astrophysical Observatory of the Russian Academy of Sciences (SAO RAS)
website at \url{http://www.sao.ru/blcat/}.

\begin{acknowledgements}

We are very grateful to other online astronomical databases, that we actively used.
We acknowledge the NASA Extragalactic Database (NED) (\url{http://ned.ipac.caltech.edu/}),
ASDC web Tools (particularly ``SED Builde'' tool \url{http://tools.asdc.asi.it/SED}),
and Roma-BZCAT - Multi-frequency Blazar Catalogue (\url{http://www.asdc.asi.it/bzcat/}).

The RATAN-600 observations were carried out with the
financial support of the Ministry of Education and Science of the Russian
Federation (14.518.11.7054) and Russian Foundation for Basic Research (12-02-31649).
\end{acknowledgements}

\bibliographystyle{mn2e} 
\bibliography{blcat} 

\label{lastpage}

\end{document}